\documentclass[a4paper,11pt]{article}
\usepackage{pos}
\usepackage{lineno}

\title{The simulated performance of GRANDProto300}
 \ShortTitle{The simulated performance of GRANDProto300}

\author*[a]{Kai-Kai Duan}
\author[a]{Peng-Xiong Ma}
\author[a]{Ke-Wen Zhang}
\author[a]{Xiao-Yuan Huang}
\author[a]{Yi Zhang}

\affiliation[a]{Key Laboratory of Dark Matter and Space Astronomy, Purple Mountain Observatory, Chinese Academy
of Sciences \\ 
No. 10 Yuanhua Road, Nanjing, China}

\onbehalf{for the GRAND collaboration} 

\emailAdd{duankk@pmo.ac.cn}

\abstract{
\textcolor{black}{GRANDProto300 is a 300-antenna prototype array of the envisioned GRAND (Giant Radio Array for Neutrino Detection) project. The goal of GRANDProto300 is to detect radio signals emitted by cosmic ray-induced air showers, with energies ranging from $10^{16.5}$~eV to $10^{18.5}$~eV, which covers the transition region between Galactic and extragalactic sources. We use simulations to optimize the layout of GRANDProto300 and develop a shower reconstruction method. Based on them, we present the performance of GRANDProto300 for cosmic-ray detection, by means of its effective area, angular resolution, and energy resolution.}
}

\ConferenceLogo{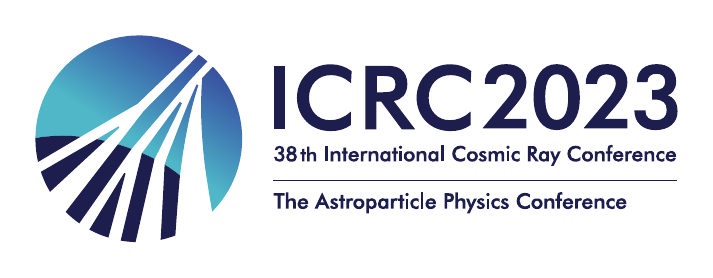}

\FullConference{%
38th International Cosmic Ray Conference (ICRC2023)\\
  26 July - 3 August, 2023\\
  Nagoya, Japan}

\begin{document}
\maketitle
% \linenumbers

\section{Introduction}

The origin of ultra-high-energy cosmic rays (UHECRs) remains a longstanding mystery in \textcolor{black}{astrophysics~\cite{2011ARA&A..49..119K}}. 
Despite their observation spanning several decades, the precise origins of UHECRs have yet to be definitively determined.
To address this challenge, \textcolor{black}{the Giant} Radio Array for Neutrino Detection (GRAND) \textcolor{black}{is an envisioned} large-scale observatory dedicated to the discovery and study of UHECR \textcolor{black}{sources~\cite{2020SCPMA..6319501A}}.
\textcolor{black}{The objectives of GRAND} encompass the \textcolor{black}{detection} and analysis of UHE cosmic rays, gamma rays, and \textcolor{black}{neutrinos, with energies above $10^{17}$~eV}.
When UHE cosmic rays, gamma rays, and neutrinos reach the Earth, the extensive air showers they induce in the atmosphere interact with the geomagnetic field, generating radio emissions that can be detected by antennas \textcolor{black}{positioned far from} the shower.

The detection of extensive air showers using radio signals has been successfully demonstrated by various experiments, including AERA, CODALEMA, LOFAR, and Tunka-Rex, as \textcolor{black}{detailed in~\cite{2016PhR...620....1H, 2017PrPNP..93....1S, 2017PTEP.2017lA106H}}.
The characteristics of radio emission, such as amplitude, frequency, and polarization, have been extensively studied and can be accurately modeled.
% Conversely, the radio-detection of an air shower by antennas at different locations can be used to infer the properties of the primary particles.
GRANDProto300 serves as a 300-antenna prototype array for the larger GRAND project. Operating within the \textcolor{black}{50--200} MHz frequency range, GRANDProto300 is specifically designed to detect inclined cosmic rays.
By leveraging the simulation results, we gain insights into the capabilities of GRANDProto300 in accurately detecting and characterizing cosmic-ray events.
This evaluation provides valuable information for the advancement of the GRAND project and its ultimate goal of understanding the origins of UHECRs.

\section{Simulation Data}

We performed simulations using ZHAireS 1.0.30a\footnote{http://aires.fisica.unlp.edu.ar/zhaires}\textcolor{black}{~\cite{2012APh....35..325A}}, a specialized tool for simulating extensive air showers and the radio emission induced by UHECRs.
Our simulations involved 16,000 UHE protons \textcolor{black}{entering the atmosphere} located at Dunhuang \textcolor{black}{in the Gansu} Province, China.
The primary protons had energies ranging from 60 PeV to 4 EeV, zenith angles between 54$^{\circ}$ and 84$^{\circ}$, and azimuth angles between 0$^{\circ}$ and 360$^{\circ}$.

\begin{figure}[!h]
    \centering
    \includegraphics[width=0.3\textwidth]{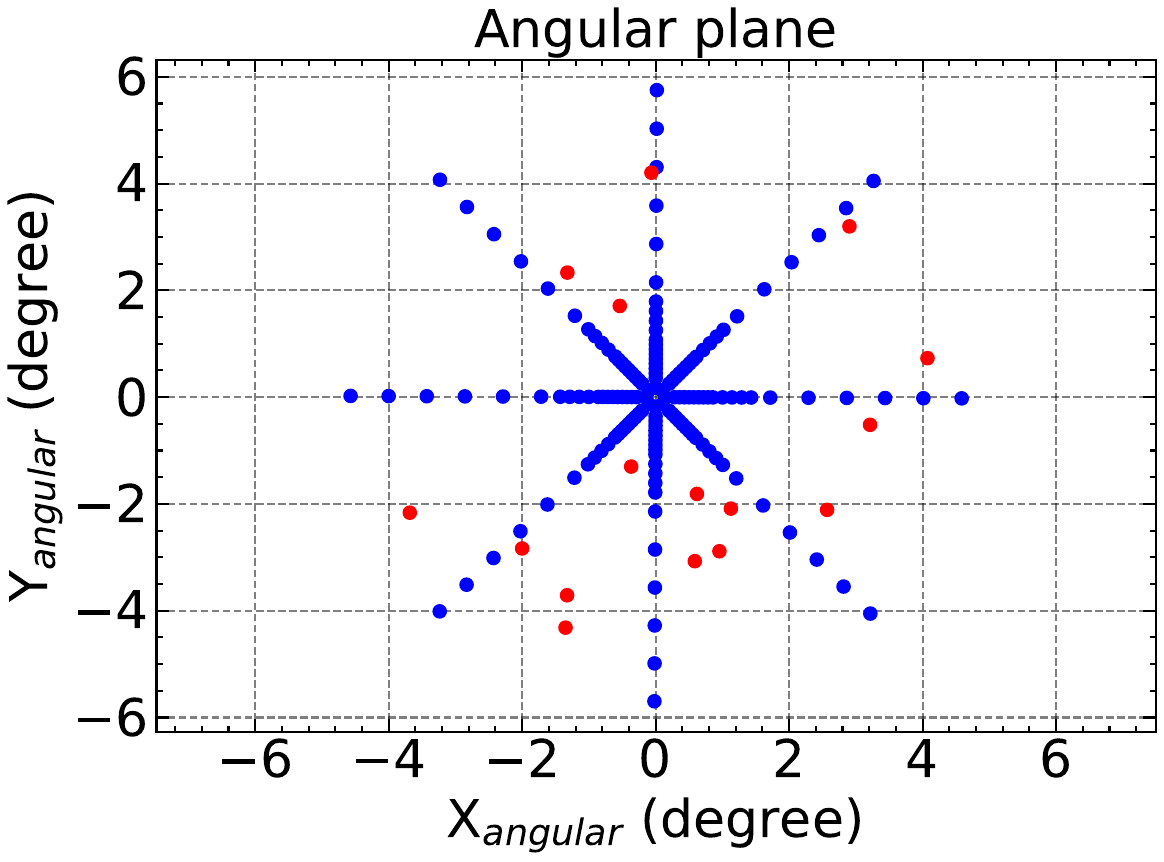}
    \includegraphics[width=0.3\textwidth]{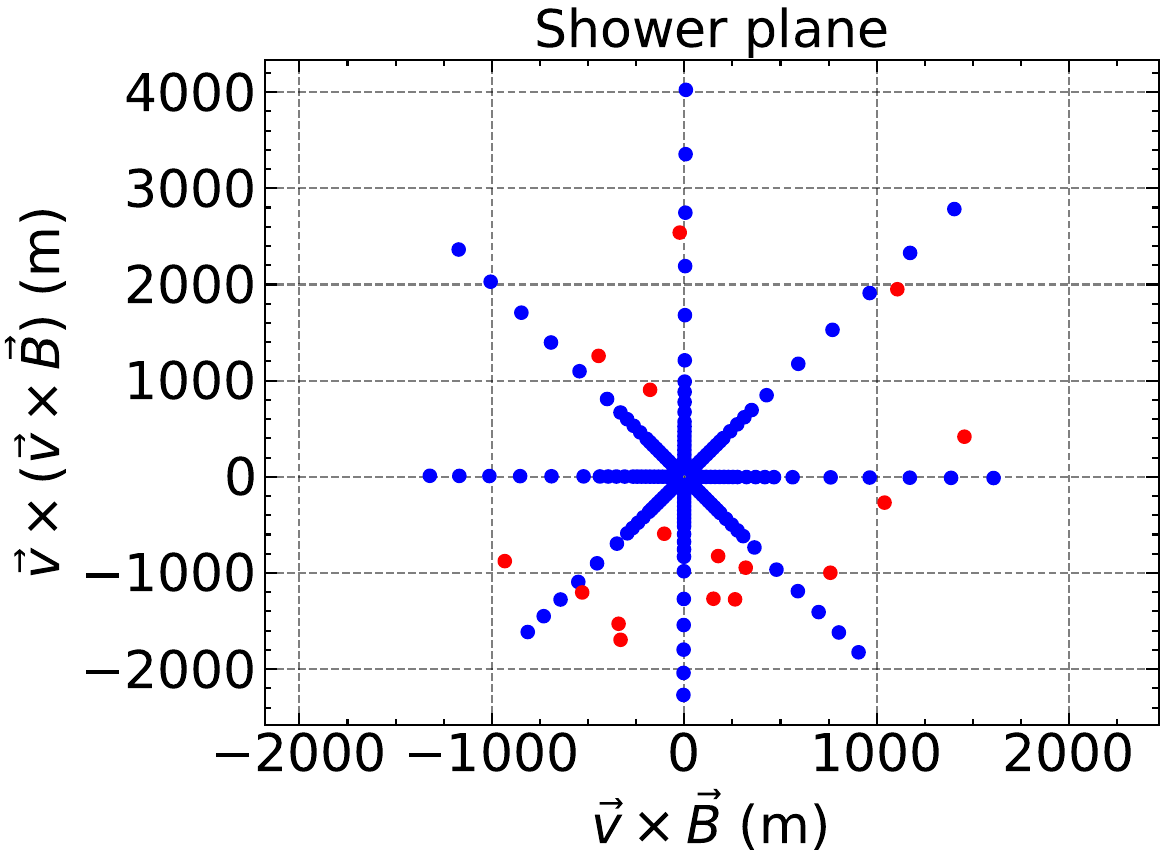}
    \includegraphics[width=0.3\textwidth]{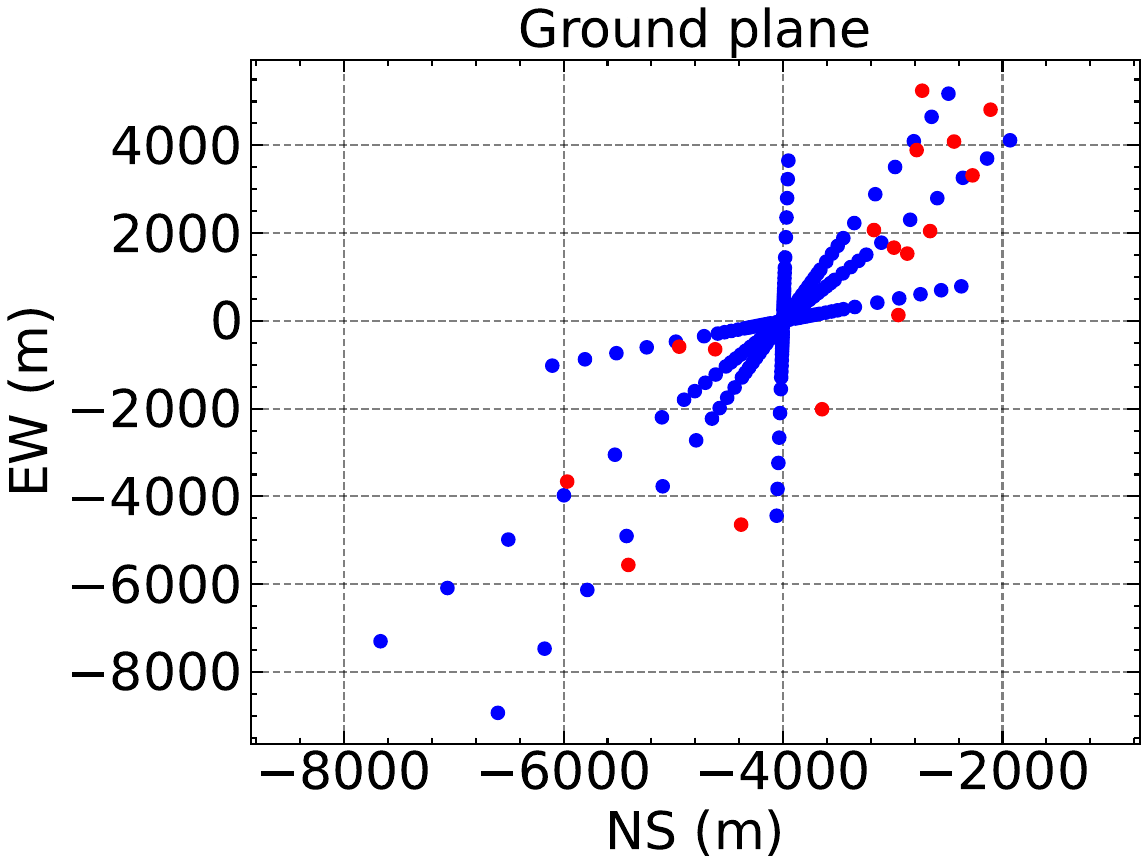}
    \caption{An example of the star-shaped layout in the angular plane (left panel), shower plane (central panel) and ground plane (right panel), simulated using proton event (with energy 4~EeV, zenith angle 71$^{\circ}$ and azimuth angle 45$^{\circ}$). \textcolor{black}{In the central panel, $\vec{v}$ and $\vec{B}$ are the directions of the shower axis and geomagnetic field, respectively.} The 160 blue points represent the antennas used for interpolation, while the 16 red points indicate the antennas used for verification.}
    \label{fig:starShapelayout}
\end{figure}

For the simulation, we utilized a \textcolor{black}{``star-shaped''} array layout, comprising 160 antennas distributed across 8 arms for \textcolor{black}{interpolation of the arrival time and the peak value of the radio emission signal} at each antenna position and 16 antennas for verification.
This configuration allows for complete coverage of the radio emission patterns, equivalent to 8 times \textcolor{black}{the diameter of the Cherenkov ring}\footnote{\textcolor{black}{For an observer located at this specific angle to
the shower, the radio signals emitted from all points along
the shower arrive simultaneously, boosting the signal along a
“Cherenkov ring”.}}.
The layout was then projected onto the ground plane based on the direction of the \textcolor{black}{cosmic-ray} air shower. 
Figure~\ref{fig:starShapelayout} illustrates an example of the antenna \textcolor{black}{layout} in the \textcolor{black}{angular plane which displays the angular positions of the antenna with respect to the \textcolor{black}{shower axis; in the shower plane, defined} as the projection of \textcolor{black}{the position of the antennas} into a plane perpendicular to the shower axis and set at the shower core \textcolor{black}{location; and in the ground plane, where} the antennas are placed}.

\begin{figure}[!h]
    \centering
    \includegraphics[width=0.4\textwidth]{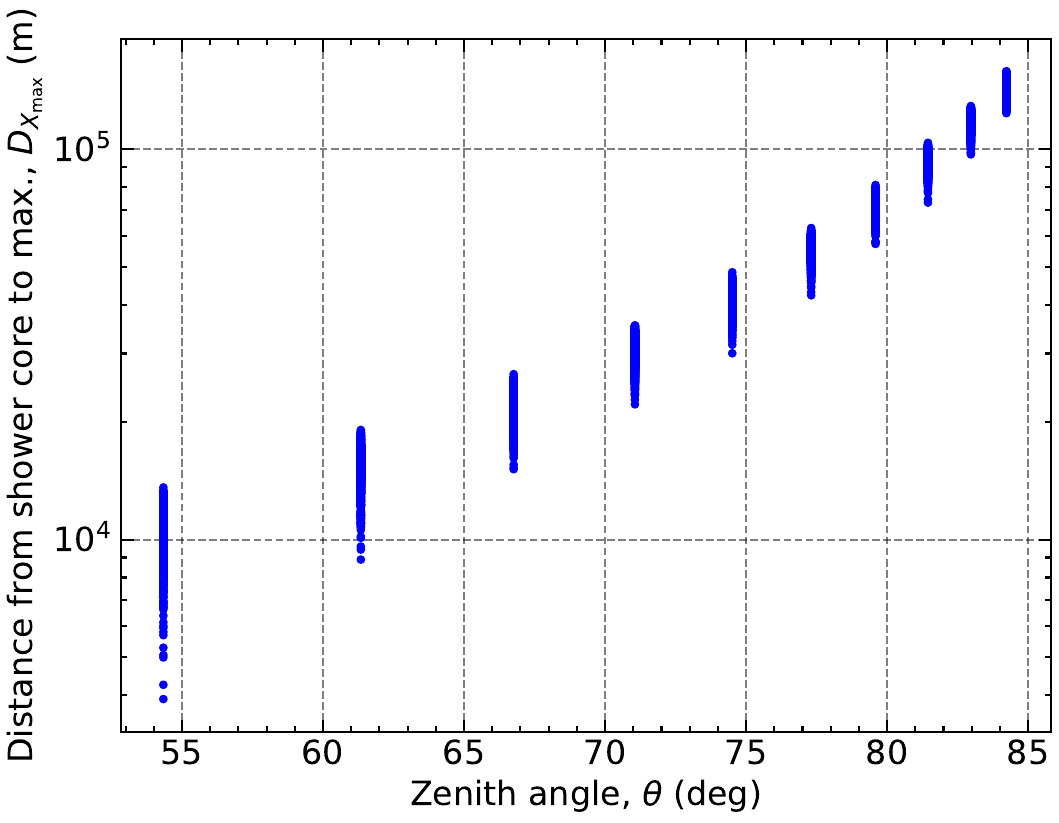}
    \includegraphics[width=0.4\textwidth]{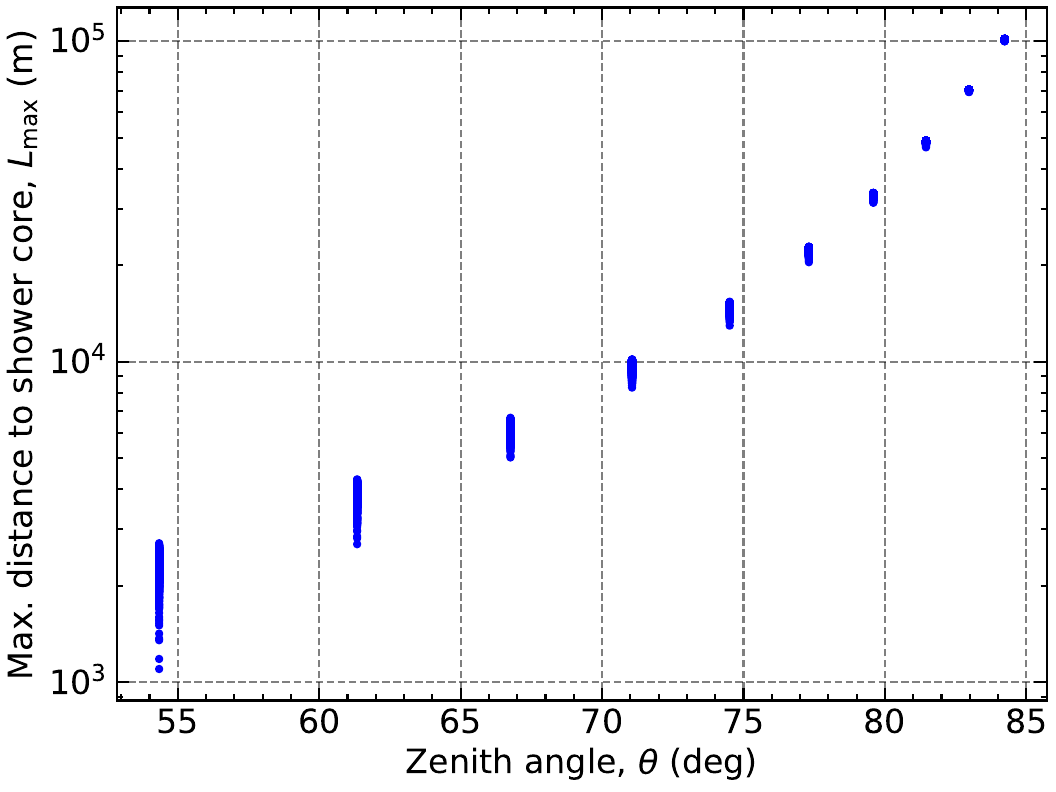}
    \caption{\textcolor{black}{\it Left:} Distances between the shower maximum and the shower core as a function of the zenith angle of the particles. \textcolor{black}{\it Right:} Maximum footprint distances between the antennas and the shower core. The dispersion observed at each zenith angle is a result of the fluctuation of the shower maximum position caused by variations in the energy of the cosmic rays and their interactions with the atmosphere.}
    \label{fig:maxL_Z}
\end{figure}

The \textcolor{black}{maximum distance, $L_{\rm max}$, between} the antennas and \textcolor{black}{shower core on the ground plane} is primarily determined by the distances from shower core to shower maximum (\textcolor{black}{$D_{X_{\rm max}}$}), the Cherenkov angle ($\omega_c = \arccos \frac{1}{n}$, where $n$ is the refraction index at the shower maximum) and the zenith angle ($\theta$) of the incoming particles:
\textcolor{black}{$L_{\rm max} = 8 D_{X_{\rm max}} \omega_c / \cos(\theta).$} Due to the variation in the position of the shower maximum caused by the energy of the primary particles and their interactions with the atmosphere, the maximum distance between the antennas and the shower core will fluctuate even for showers with the same zenith angle.
Figure~\ref{fig:maxL_Z} illustrates the \textcolor{black}{relation} between the distances from the shower maximum to the shower core and the \textcolor{black}{maximum distances} between the antennas and the shower core, considering different zenith angles of the incoming particles.
These maximum distances play a crucial role in determining the allowable range of positions for the shower core beyond the layout of GRANDProto300.

\begin{figure}[!ht]
    \centering
    \includegraphics[width=0.3\textwidth]{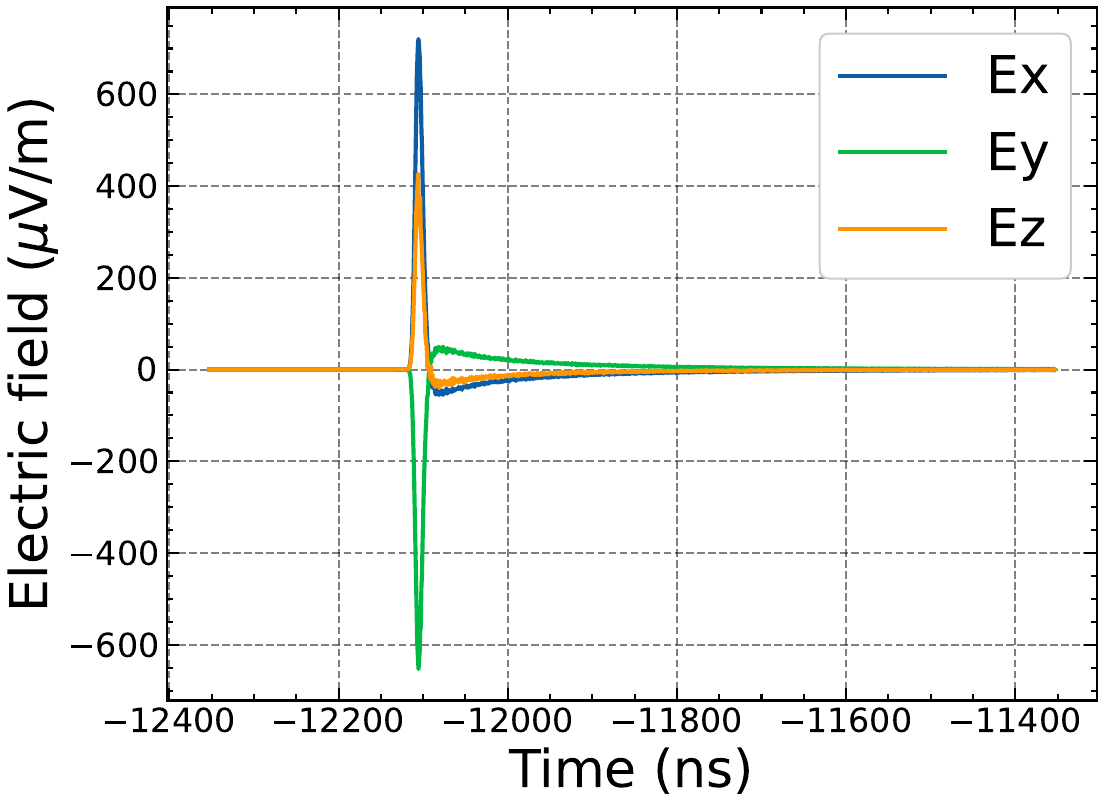}
    \includegraphics[width=0.3\textwidth]{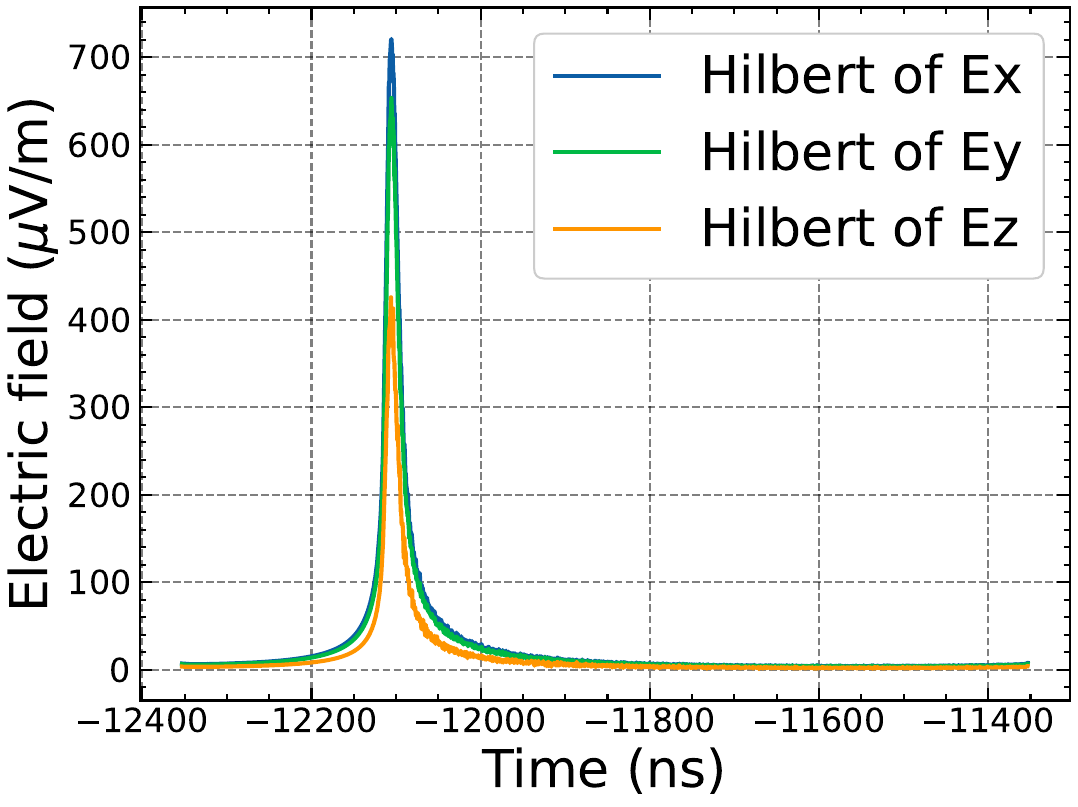}
    \includegraphics[width=0.3\textwidth]{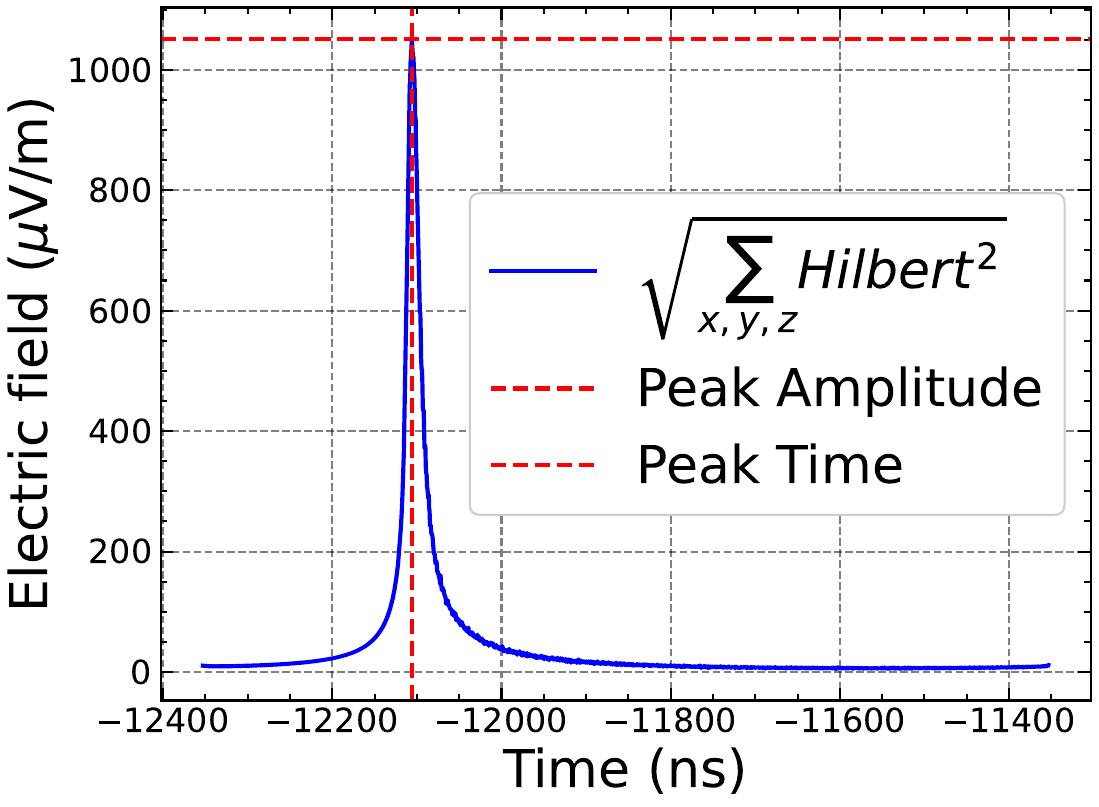}
    \includegraphics[width=0.3\textwidth]{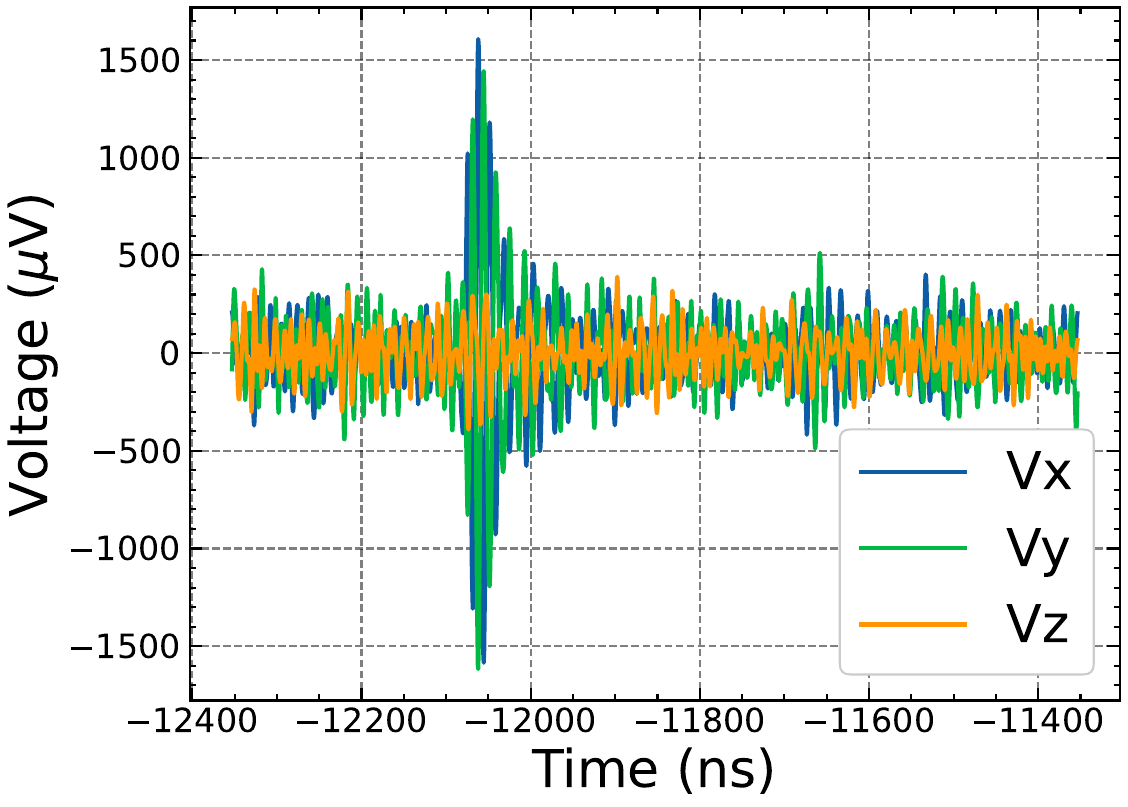}
    \includegraphics[width=0.3\textwidth]{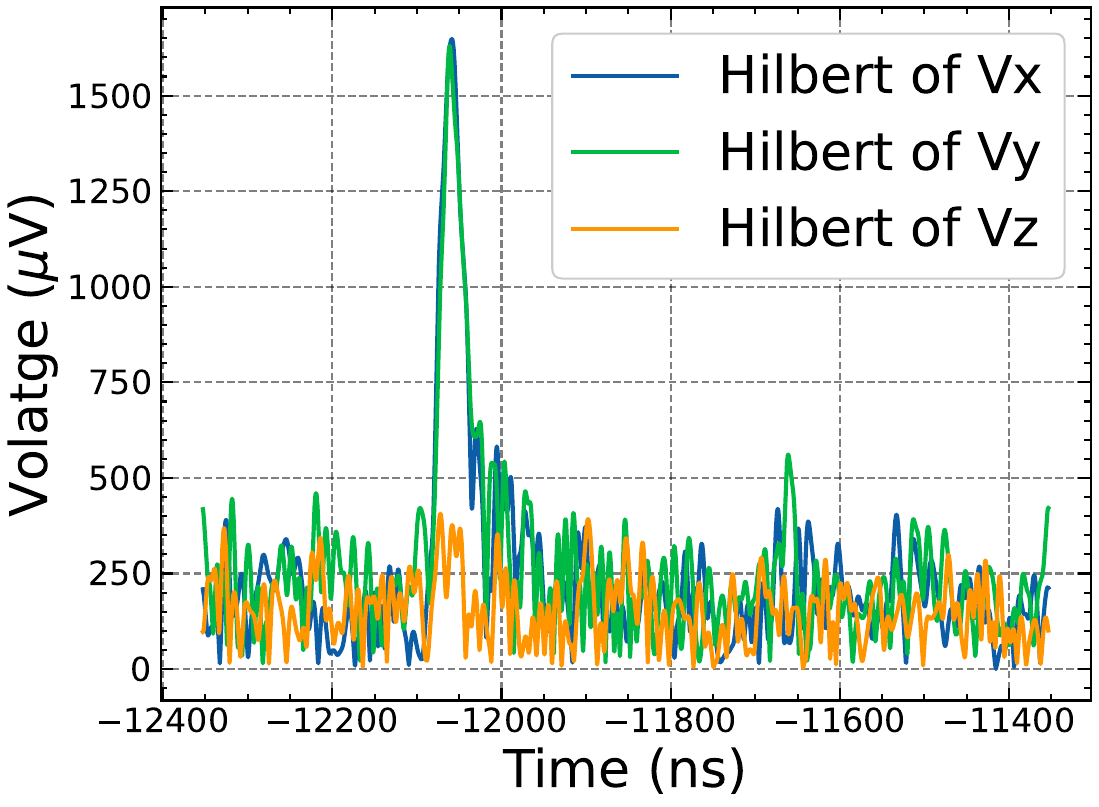}
    \includegraphics[width=0.3\textwidth]{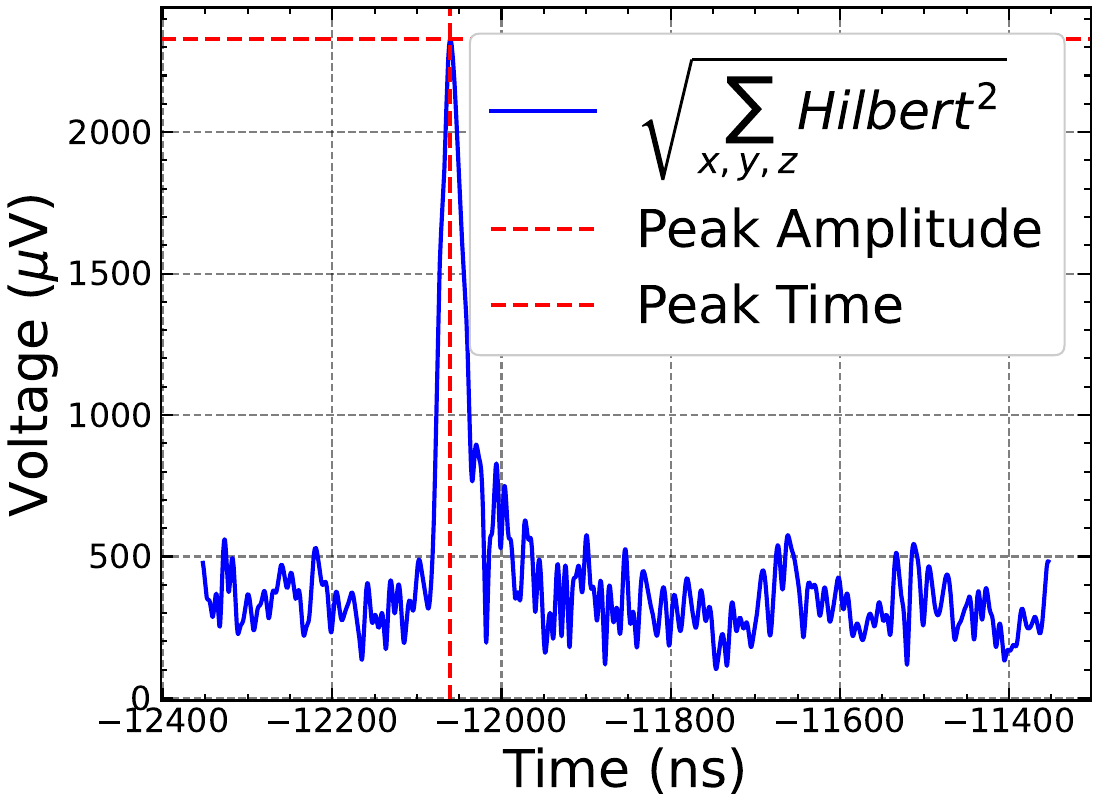}
    \caption{An example illustrating the electric fields (top row) and voltages (bottom row) of each arm (left column) of the antennas. The central column displays the Hilbert envelopes of the electric fields and voltages for each arm. The \textcolor{black}{right} column represents \textcolor{black}{the root sum of the square of the Hilbert envelope of each arm} for the electric field and voltage. These measurements were obtained from one of the antennas.}
    \label{fig:signal}
\end{figure}

After performing the ZHAireS simulation, the electric \textcolor{black}{field} at each antenna position \textcolor{black}{is} obtained.
Subsequently, we simulate the antenna response to these electric fields, resulting in the calculation of voltages for each arm of the antennas. Additionally, the voltage calculations \textcolor{black}{include the noise} from the galactic plane at 18 \textcolor{black}{hs} (LST).
Figure~\ref{fig:signal} presents an example displaying the electric fields and voltages of each arm, as well as the root sum of the Hilbert envelope of the electric field and voltage, obtained from one of the antennas.
The key characteristics of the signal are the peak time and peak amplitude observed in each arm.
These features play a significant role in the reconstruction of the primary properties of the cosmic rays.

\section{Interpolation and Reconstruction Method}

\subsection{Interpolation}

In order to assess the performance of GRANDProto300, we employ interpolation techniques to map the simulation data onto \textcolor{black}{studied} GRANDProto300 layouts, allowing us to reconstruct the properties of the primary event.

\begin{figure}[!ht]
    \centering
    \includegraphics[width=0.45\textwidth]{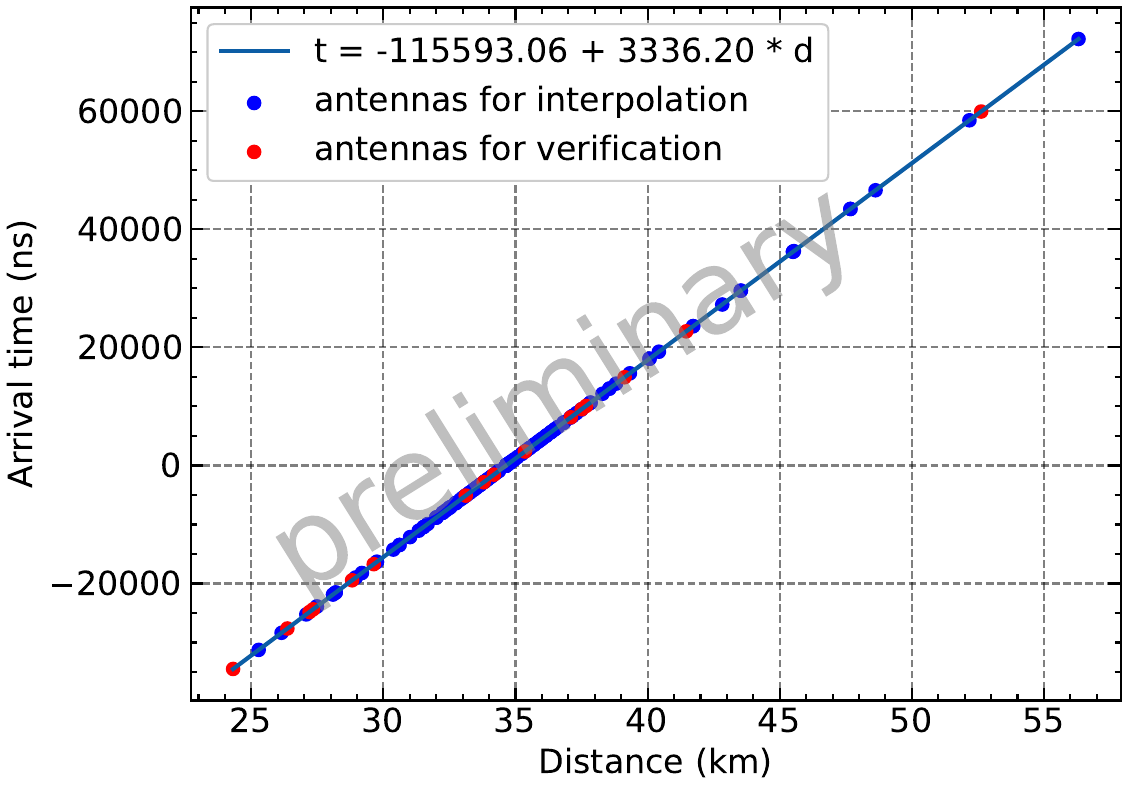}
    \includegraphics[width=0.42\textwidth]{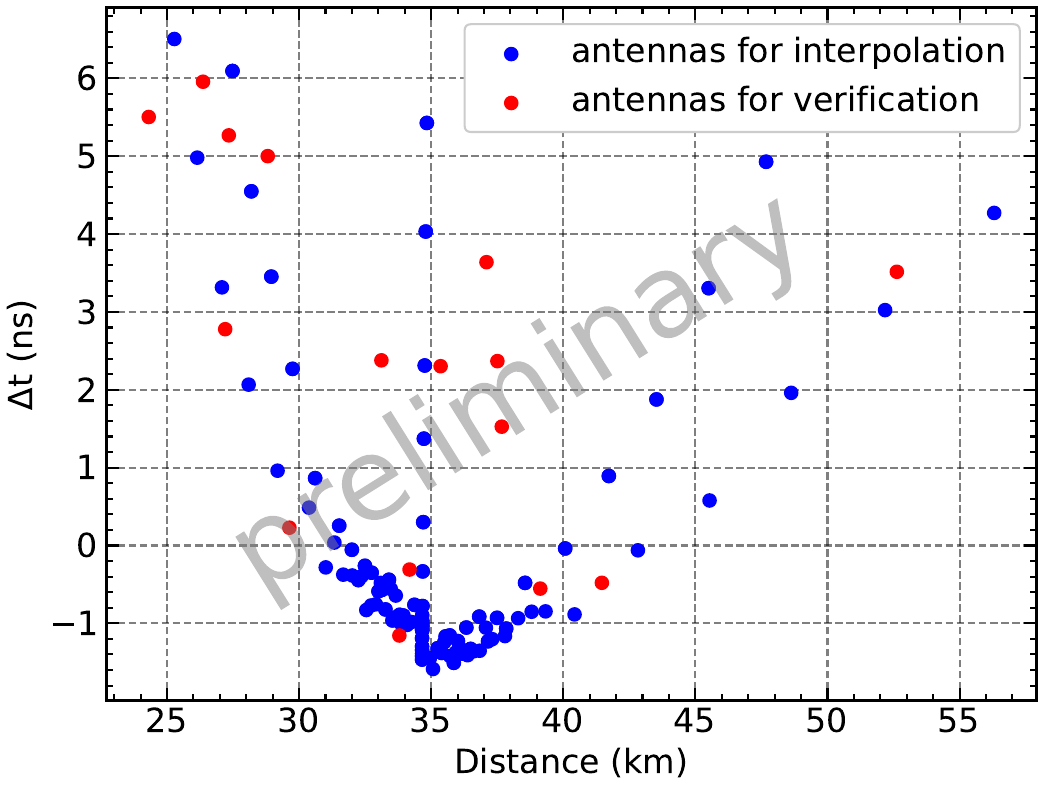}
    \caption{\textcolor{black}{\it Left:} Linear interpolation of the arrival times of each antenna as a function of the distances between the antennas and the shower maximum. \textcolor{black}{\it Right:} Residual times of antennas.}
    \label{fig:t_d}
\end{figure}

The arrival times of the signals received by each antenna exhibit a linear dependence on the distances between the antennas and the shower maximum. 
Figure~\ref{fig:t_d} illustrates the linear interpolation of the arrival times for each antenna as a function of the distances to the shower maximum.
\textcolor{black}{The} maximum residual time is found to be less than 2~ns close to the shower core, which is comparable to the time uncertainty introduced by the GPS module.

\begin{figure}[!ht]
    \centering
    \includegraphics[width=0.9\textwidth]{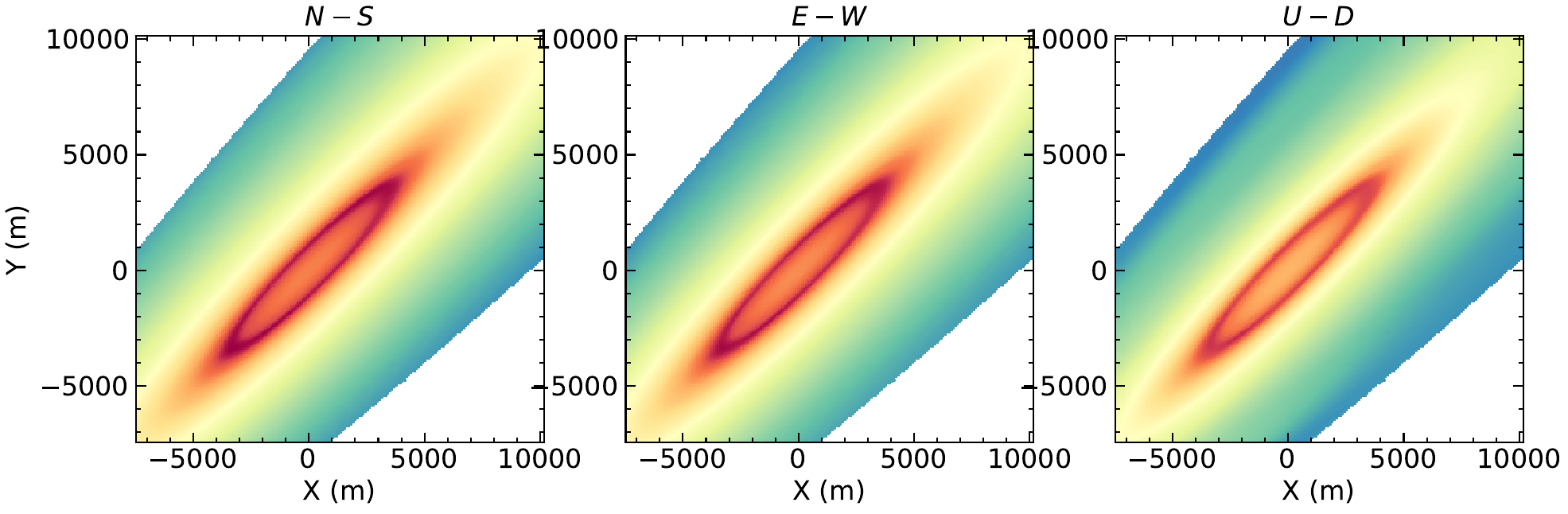}
    \caption{Interpolated results of the simulation voltage peak amplitudes for each arm \textcolor{black}{of} the star-shaped layout.}
    \label{fig:interpolation}
\end{figure}

The interpolation method for the amplitude is based on the algorithm introduced \textcolor{black}{in ~\cite{2023arXiv230613514C}}.
This method involves decomposing the radio emission into sub-components using Fourier transformation around the shower axis.
The Fourier coefficients are then interpolated based on the offset angle from the shower axis, and the composite signal is reconstructed using inverse Fourier transformation.
Figure~\ref{fig:interpolation} demonstrates the interpolated results of the simulation voltage peak amplitudes for each arm of the star-shaped layout.

\subsection{Reconstruction}

The reconstruction method utilized in this study is based on the research conducted \textcolor{black}{in~\cite{decoene:tel-03153273, Decoene:2021jF}}.
For the inclined air shower, the geometrical configurations satisfy the Fraunhofer conditions.
% $S / L \lambda \ll 1$
% where $S$ is the surface of the the emitting region $S \sim$ 10~m, $L$ is the distance between the observer and the emitting region $L \sim$ 20 - 400~km, $\lambda$ is the wavelength at which the observation is done $\lambda \sim$ 1.5-6~m (corresponding to 200~MHz and 50~MHz).
As a result, the radiation wavefront can be approximated as a plane wave at first order, allowing for a rough reconstruction of the emission direction.
The study of the wavefront indicates that the spherical model provides sufficient accuracy in describing the radio emission wavefront in GRAND~\cite{decoene:tel-03153273}.
Although the spherical model does not enable the direct reconstruction of the arrival direction of the emission, it allows for the identification of the emission source as a point-like source.
To reconstruct the arrival direction, the shower core position is required as the second point along the shower axis, which can be determined using the amplitude distribution.

The reconstruction process consists of three steps. Firstly, a plane wave reconstruction technique is employed to reduce the parameter space, narrowing down the possible solutions. This step helps to establish an initial estimation of the arrival direcion.

Next, the spherical reconstruction method is applied to determine the precise position of the emission source. The spherical model takes into account the spherical wavefront characteristics and allows for a more accurate determination of the emission source position.

Finally, the amplitude profile is fitted using an Angular Distribution Function (ADF) model. This step involves matching the observed amplitude distribution with the expected behavior predicted by the ADF model. The fitting process helps to refine the reconstruction and obtain a more precise estimation of the arrival direction of the emission.

The complete reconstruction chain enables us to accurately reconstruct both the arrival direction and the emission point of the shower.
% The translation from the emission point position reconstruction to the terms of $X_{\text{max}}$ reconstruction occurs naturally, facilitating the determination of the depth of the shower maximum ($X_{\text{max}}$) within the atmosphere.
Furthermore, an important parameter in the ADF reconstruction, denoted as $\mathcal{A}$, is directly related to the total energy of shower emission. This relation allows us to utilize $\mathcal{A}$ as a tool to reconstruct the primary energy of cosmic rays. By analyzing the value of $\mathcal{A}$ obtained through the ADF fitting, we can obtain valuable information about the energy of the incident cosmic rays.
The combination of reconstructing the arrival direction, emission point, and primary energy provides a comprehensive understanding of the cosmic ray events detected by GRANDProto300, enabling us to study and analyze the properties and sources of UHECRs.

\section{Simulated Performance of GRANDProto300}

By employing \textcolor{black}{interpolation}, the simulation data obtained from the star-shaped layout can be interpolated to the GRANDProto300 layout.
This interpolation is carried out by considering random shower core positions that lie within circles with a radius equal to the maximum footprint distance.
To evaluate the acceptance of the GRANDProto300 array, certain criteria are applied.
Specifically, events that meet the trigger condition are selected.
The trigger condition necessitates that the signal is detected by at least five antennas, surpassing a threshold set at 5 times the standard deviation of the \textcolor{black}{Galactic} noise.
By applying these selection criteria, the acceptance of the GRANDProto300 array can be estimated.
This estimation provides valuable insights into the array's capability to detect and capture relevant signals from the simulated events.

\begin{figure}[!ht]
    \centering
    \includegraphics[width=0.245\textwidth]{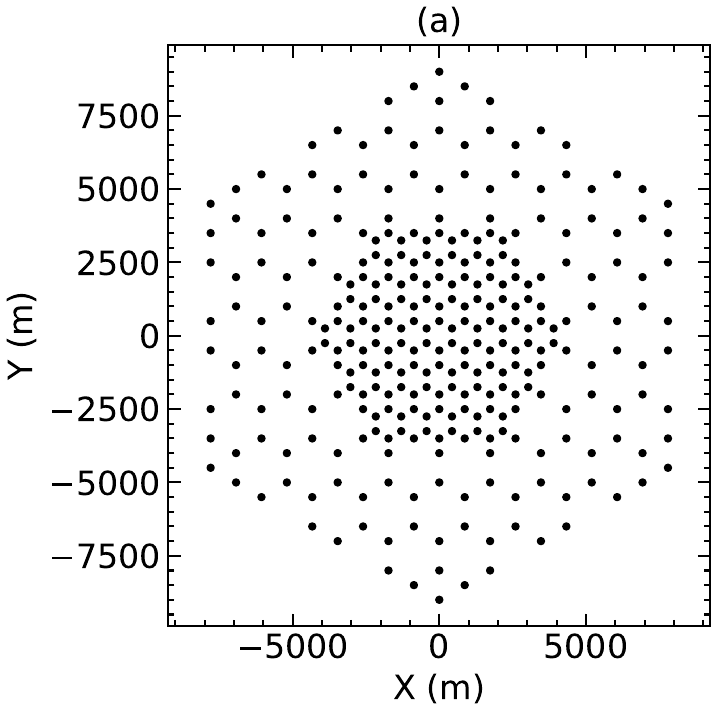}
    \includegraphics[width=0.245\textwidth]{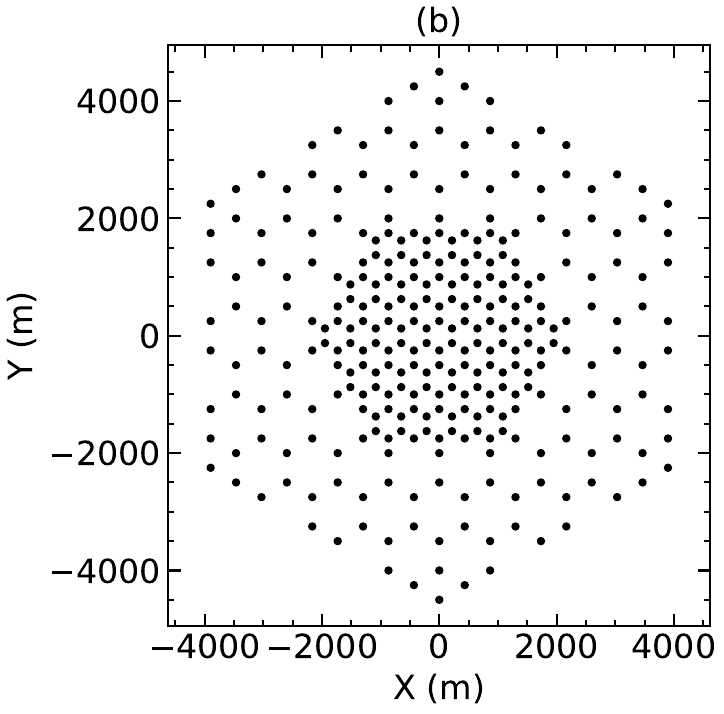}
    \includegraphics[width=0.245\textwidth]{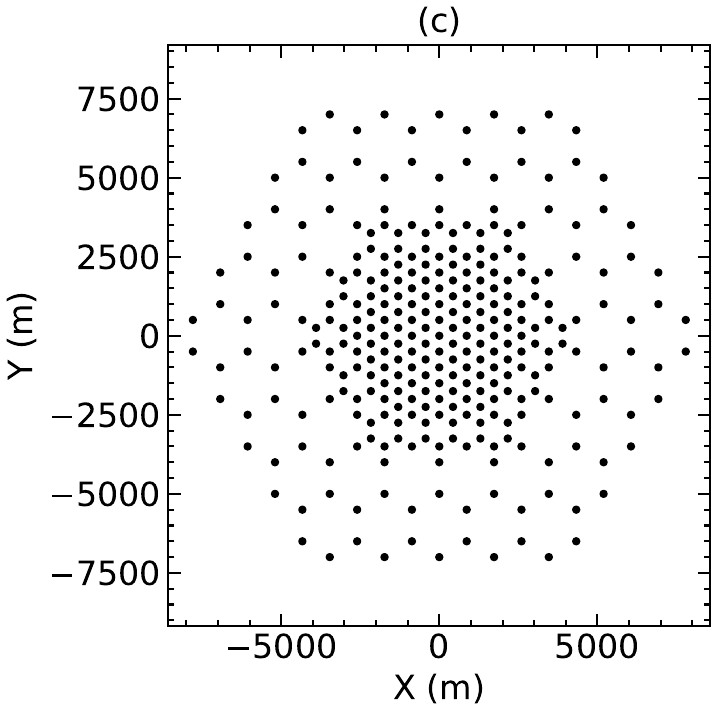}
    \includegraphics[width=0.245\textwidth]{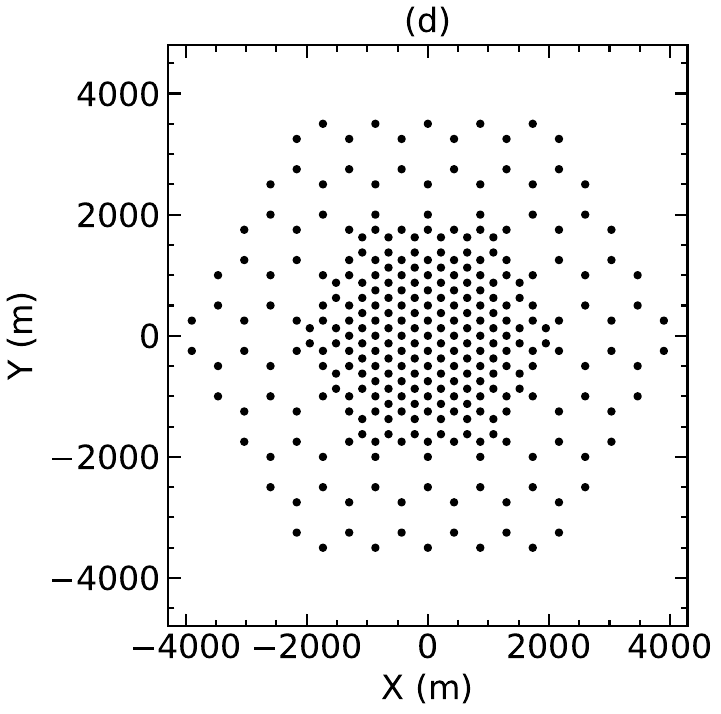}
    \caption{Four different layouts of the GRANDProto300 \textcolor{black}{under consideration}. \textcolor{black}{(a) Layout with a 1~km inter-station spacing hexagonal outer region and a 500~m inter-station spacing hexagonal inner region. (b) Layout with a 500~m inter-station spacing hexagonal outer region and a 250~m inter-station spacing hexagonal inner region. (c) Layout with a 1~km inter-station spacing hexagonal outer region and a 500~m inter-station spacing triangle inner region. (d) Layout with a 500~m inter-station spacing hexagonal outer region and a 250~m inter-station spacing triangle inner region.}}
    \label{fig:GP300layout}
\end{figure}

\begin{figure}[!ht]
    \centering
    \includegraphics[width=0.4\textwidth]{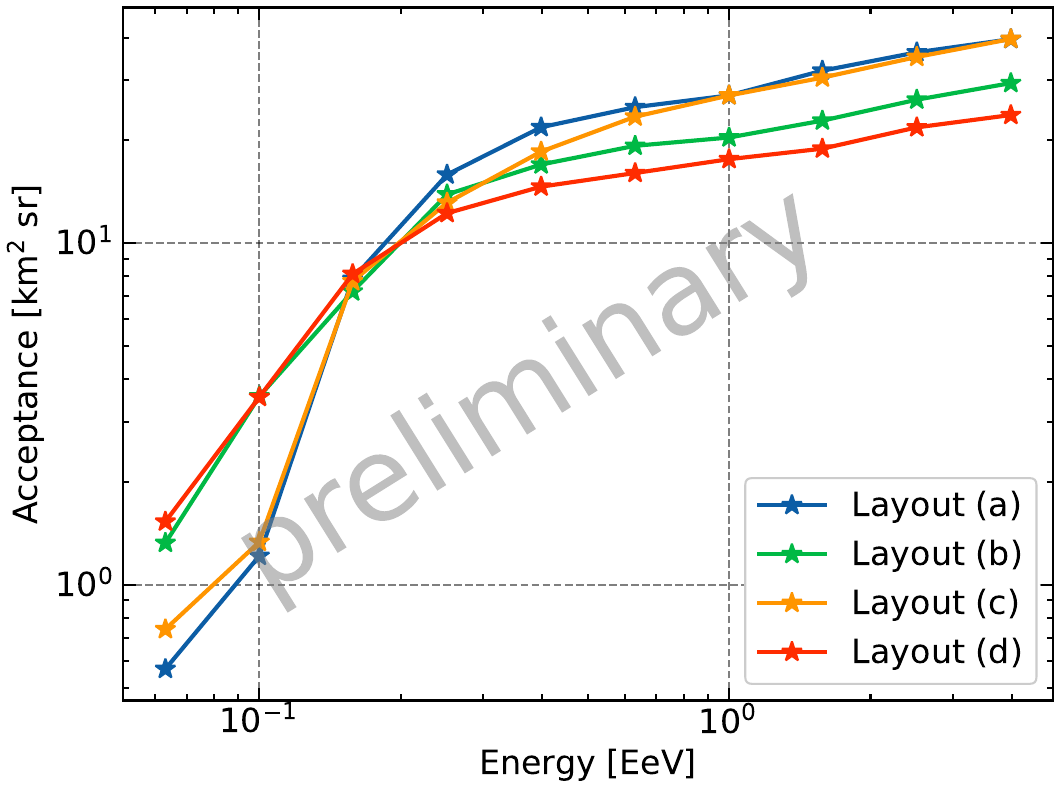}
    \includegraphics[width=0.4\textwidth]{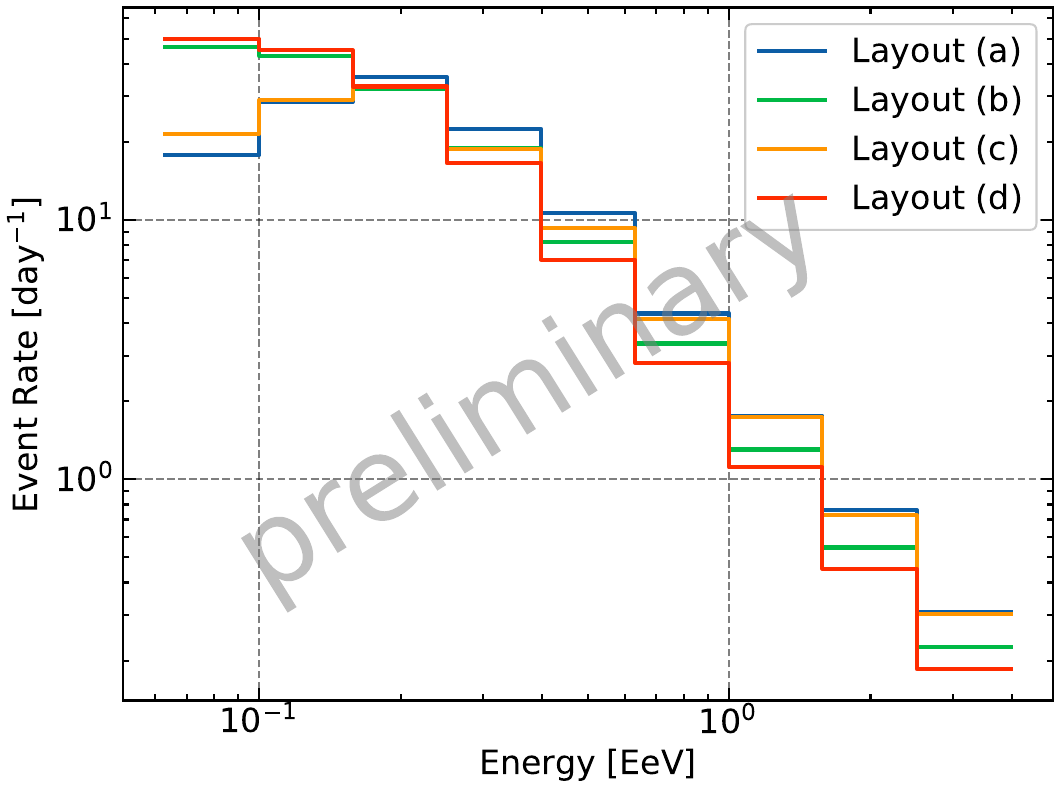}
    \caption{The expected acceptances (\textcolor{black}{\it left}) and event rates (\textcolor{black}{\it right}) for \textcolor{black}{the detection of UHE proton-initiated extensive air showers} by different GRANDProto300 layouts.}
    \label{fig:Acc}
\end{figure}

In order to optimize the layout of the GRANDProto300 array, four \textcolor{black}{competing layout options} are compared.
These layouts consist of a hexagonal outer region with 1 km and 500 m spacing, as well as a hexagonal or triangular inner region with 500 m to 250 m spacing.
\textcolor{black}{Figure}~\ref{fig:GP300layout} illustrates these four different layouts of the GRANDProto300 array.
To evaluate the performance of these layouts, the interpolation and trigger selection techniques are applied to estimate their acceptance.
By calculating the acceptance, we can assess the \textcolor{black}{capability of the array} to detect and capture relevant signals from simulated events.
Considering the proton flux observed by TALE~\cite{2018ApJ...865...74A}, we \textcolor{black}{derive} the expected event rate of protons detected by the GRANDProto300 array.
Figure~\ref{fig:Acc} illustrates the expected acceptance and event rate for proton detection across the different layouts of GRANDProto300.
By comparing these layouts, we observe that a sparse outer region combined with a dense inner region leads to an enlarged acceptance in both the high-energy and low-energy ranges.
% The comparison of these layouts provides valuable insights into the design considerations for optimizing the GRANDProto300 array, enabling researchers to enhance its performance and maximize the detection capabilities for protons.

\begin{figure}[!ht]
    \centering
    \includegraphics[width=0.4\textwidth]{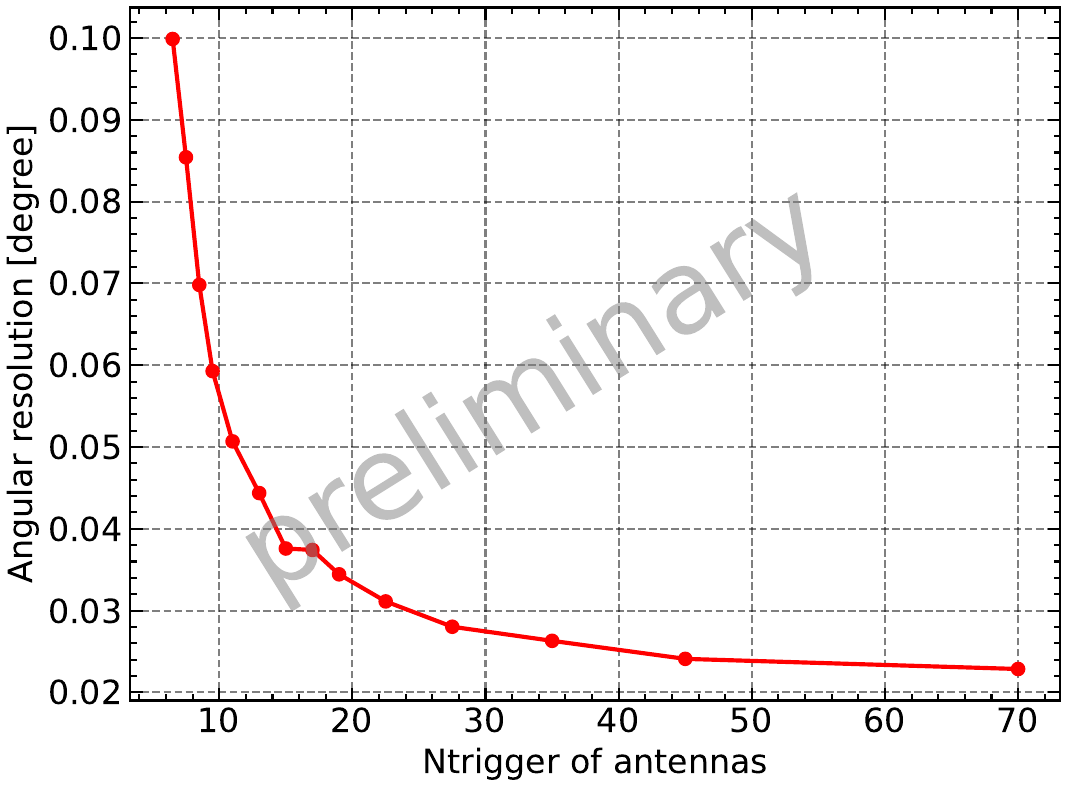}
    \includegraphics[width=0.4\textwidth]{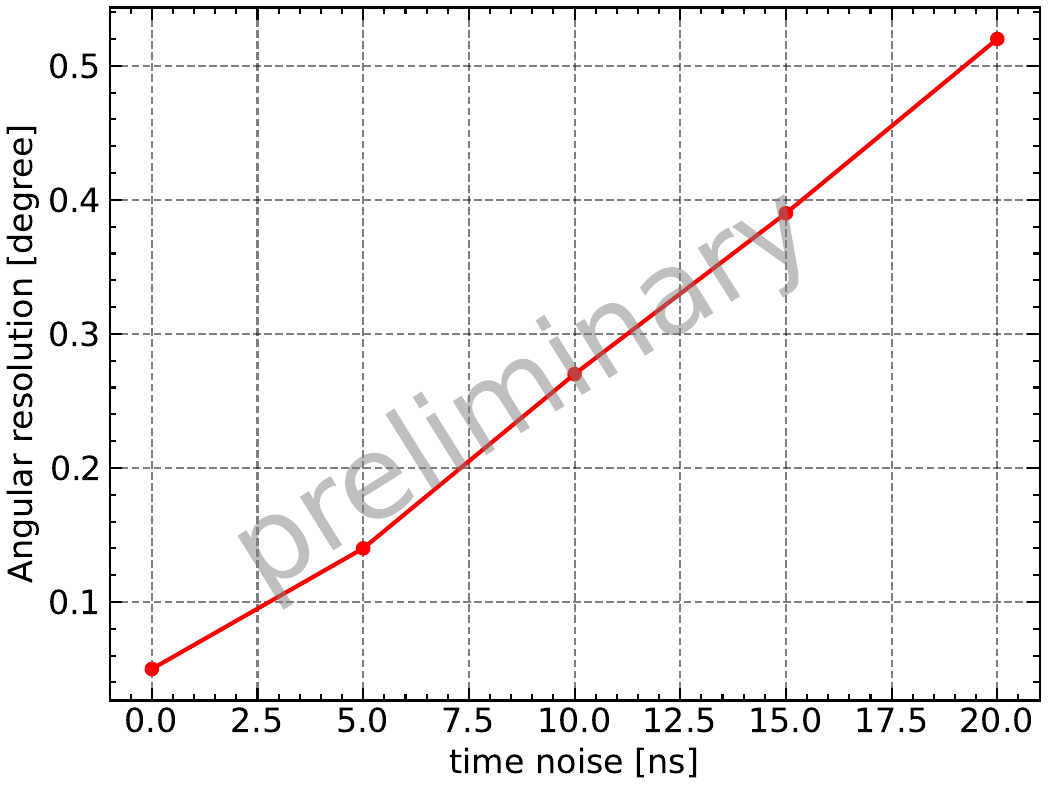}
    \caption{\textcolor{black}{\it Left:} \textcolor{black}{Variation of the angular resolution of GRANDProto300 with the number of triggered antennas.} As the number of triggered antennas increases, the angular resolution improves. \textcolor{black}{\it Right:} Dependence of the angular resolution on the time noise of the trigger time. With lower time noise, the angular resolution of \textcolor{black}{GRANDProto300} is enhanced.}
    \label{fig:dAngle}
\end{figure}

The angular resolution of GRANDProto300 is defined as the \textcolor{black}{range of angles} containing 68$\%$ of the distribution of the angular distance between the reconstructed direction and the primary direction of the shower.
% given by:
% $$
% \cos(\mathrm d \theta) = \cos(\theta_{\text{rec}}) \cos(\theta_{\text{pri}}) + \cos(\phi_{\text{rec}} - \phi_{\text{pri}}) \sin(\theta_{\text{rec}}) \sin(\theta_{\text{pri}}),
% $$
% where $\mathrm d \theta$ is the angular distance, $\theta_{\text{rec}}$ and $\phi_{\text{rec}}$ is the reconstructed zenith and azimuth angle, $\theta_{\text{pri}}$ and $\phi_{\text{pri}}$ is the primary zenith and azimuth angle.
Figure~\ref{fig:dAngle} shows \textcolor{black}{how} the angular resolution of GRANDProto300 varies \textcolor{black}{with} number of the triggered antennas and the noise of the trigger time.
 It can be observed that the angular resolution of GRANDProto300 improves as the number of the triggered antennas increase, while it worsens with increased noise in the trigger time.
Overall, the angular resolution of GRANDProto300 is better than 0.3 degrees when the time noise is within 10 ns.

\begin{figure}[!ht]
    \centering
    \includegraphics[width=0.4\textwidth]{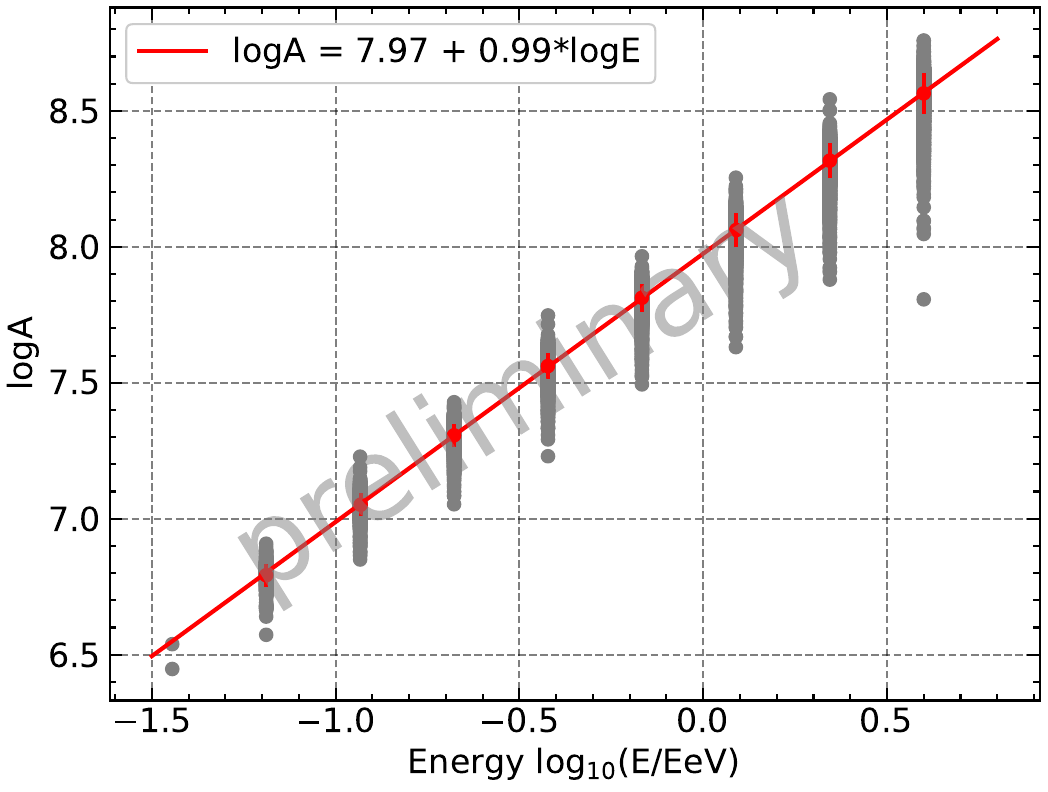}
    \includegraphics[width=0.4\textwidth]{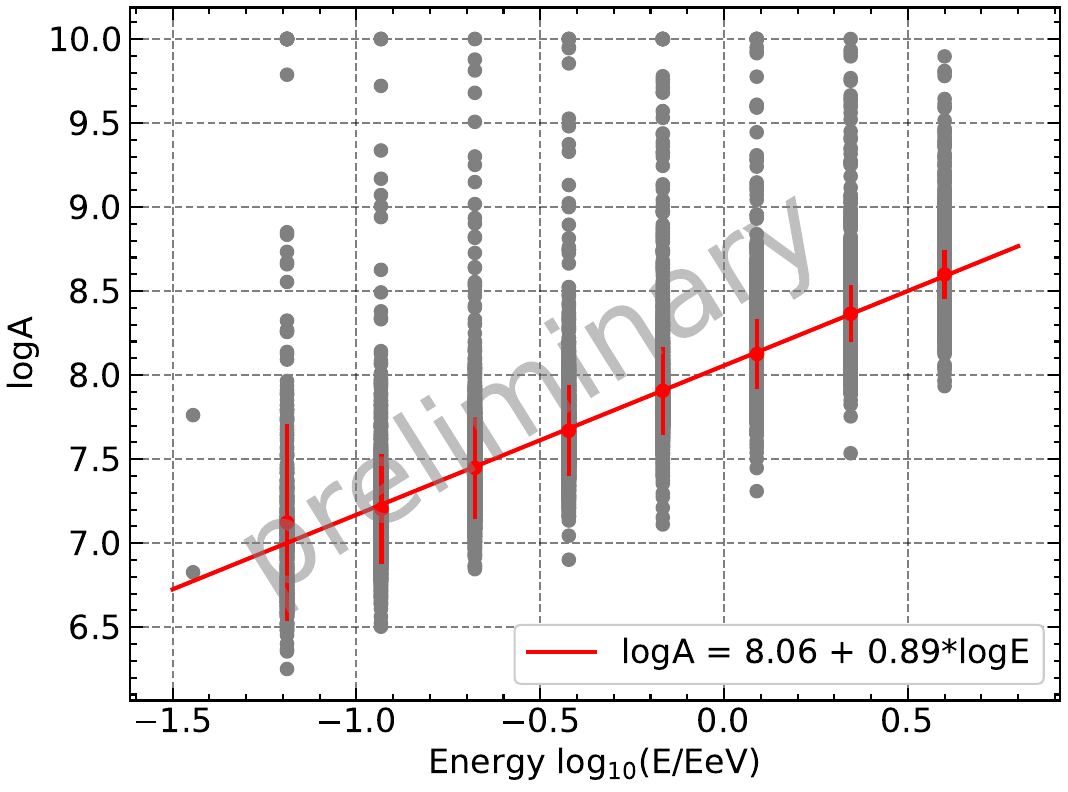}
    \caption{\textcolor{black}{\it Left:} Linear \textcolor{black}{relation} between the parameter $\mathcal{A}$ and the primary energy of the shower without time noise. \textcolor{black}{\it Right:} Same \textcolor{black}{relation} with the inclusion of 10~ns noise in the trigger times.}
    \label{fig:energy}
\end{figure}

Figure~\ref{fig:energy} shows the linear relation between the parameter $\mathcal{A}$ and the primary energy of the shower.
The parameter $\mathcal{A}$ can be used as an indicator to estimate the primary energy of the cosmic rays detected by GRANDProto300. However, it should be noted that the energy resolution of GRANDProto300 deteriorates when considering trigger times with 10 ns noise.

\section{Summary}

\textcolor{black}{Using} simulation data, we \textcolor{black}{explore four possible layouts} of GRANDProto300 and implemented a reconstruction method based on\textcolor{black}{~\cite{decoene:tel-03153273, Decoene:2021jF}}.
By interpolating the simulation data from the star-shaped layout to the GRANDProto300 layout and applying the trigger condition, we have determined the acceptance and estimated the event rate.
Furthermore, through the process of plane wave, spherical wave, and ADF reconstruction, we have successfully reconstructed the arrival direction and \textcolor{black}{position of the emission point} of the showers.
Moreover, the parameter $\mathcal{A}$ in the ADF reconstruction is directly correlated with the primary energy of the shower emission, which allows us to reconstruct the energy of the cosmic rays accurately.

\section*{Acknowledgements}
This work is supported by National Natural Science Foundation of China under grants 12273114, the Chinese Academy of Sciences, and the Program for Innovative Talents and Entrepreneurship in Jiangsu.

\newcommand\aap{Astron. Astrophys.}
\newcommand\aj{Astron. J.}
\newcommand\ap{Astropart. Phys.}
\newcommand\apjs{Astrophys. J. Suppl. Ser.}
\newcommand\araa{Ann. Rev. Astron. Astrophys.}
\newcommand\mnras{Mon. Not. R. Astron. Soc.}
\newcommand\jcap{J. Cosmol. Astropart. Phys.}
\newcommand\ssr{Space Sci. Rev.}
\newcommand\pasj{Publ. Astron. Soc. Jpn.}
\newcommand\aaps{Astron. Astrophys. Suppl. Ser.}
\newcommand\pasa{Publ. Astron. Soc. Aust.}
\newcommand\apj{Astrophys. J.}
\newcommand\apjl{Astrophys. J. Lett.}
\newcommand\physrep{Phys. Rep.}
\newcommand\raa{Res. Astron. Astrophys.}
\newcommand\prl{Phys. Rev. Lett.}
\newcommand\prd{Phys. Rev. D}

\bibliographystyle{apsrev}
\setlength{\bibsep}{0.1ex}
\bibliography{skeleton}

\begin{thebibliography}{10}
\expandafter\ifx\csname natexlab\endcsname\relax\def\natexlab#1{#1}\fi
\expandafter\ifx\csname bibnamefont\endcsname\relax
  \def\bibnamefont#1{#1}\fi
\expandafter\ifx\csname bibfnamefont\endcsname\relax
  \def\bibfnamefont#1{#1}\fi
\expandafter\ifx\csname citenamefont\endcsname\relax
  \def\citenamefont#1{#1}\fi
\expandafter\ifx\csname url\endcsname\relax
  \def\url#1{\texttt{#1}}\fi
\expandafter\ifx\csname urlprefix\endcsname\relax\def\urlprefix{URL }\fi
\providecommand{\bibinfo}[2]{#2}
\providecommand{\eprint}[2][]{\url{#2}}

\bibitem[{\citenamefont{{Kotera} and {Olinto}}(2011)}]{2011ARA&A..49..119K}
\bibinfo{author}{\bibfnamefont{K.}~\bibnamefont{{Kotera}}} \bibnamefont{and}
  \bibinfo{author}{\bibfnamefont{A.~V.} \bibnamefont{{Olinto}}},
  \bibinfo{journal}{\araa} \textbf{\bibinfo{volume}{49}}, \bibinfo{pages}{119}
  (\bibinfo{year}{2011}), \eprint{1101.4256}.

\bibitem[{\citenamefont{{{\'A}lvarez-Mu{\~n}iz}
  et~al.}(2020)}]{2020SCPMA..6319501A}
\bibinfo{author}{\bibfnamefont{J.}~\bibnamefont{{{\'A}lvarez-Mu{\~n}iz}}}
  \bibnamefont{et~al.}, \bibinfo{journal}{Sci. China Phys. Mech. Astron.}
  \textbf{\bibinfo{volume}{63}}, \bibinfo{eid}{219501} (\bibinfo{year}{2020}),
  \eprint{1810.09994}.

\bibitem[{\citenamefont{{Huege}}(2016)}]{2016PhR...620....1H}
\bibinfo{author}{\bibfnamefont{T.}~\bibnamefont{{Huege}}},
  \bibinfo{journal}{\physrep} \textbf{\bibinfo{volume}{620}},
  \bibinfo{pages}{1} (\bibinfo{year}{2016}), \eprint{1601.07426}.

\bibitem[{\citenamefont{{Schr{\"o}der}}(2017)}]{2017PrPNP..93....1S}
\bibinfo{author}{\bibfnamefont{F.~G.} \bibnamefont{{Schr{\"o}der}}},
  \bibinfo{journal}{Prog. Theor. Exp. Phys.} \textbf{\bibinfo{volume}{93}},
  \bibinfo{pages}{1} (\bibinfo{year}{2017}), \eprint{1607.08781}.

\bibitem[{\citenamefont{{Huege} and {Besson}}(2017)}]{2017PTEP.2017lA106H}
\bibinfo{author}{\bibfnamefont{T.}~\bibnamefont{{Huege}}} \bibnamefont{and}
  \bibinfo{author}{\bibfnamefont{D.}~\bibnamefont{{Besson}}},
  \bibinfo{journal}{Prog. Theor. Exp. Phys.} \textbf{\bibinfo{volume}{2017}},
  \bibinfo{eid}{12A106} (\bibinfo{year}{2017}), \eprint{1701.02987}.

\bibitem[{\citenamefont{{{\'A}lvarez-Mu{\~n}iz}
  et~al.}(2012)\citenamefont{{{\'A}lvarez-Mu{\~n}iz}, {Carvalho}, and
  {Zas}}}]{2012APh....35..325A}
\bibinfo{author}{\bibfnamefont{J.}~\bibnamefont{{{\'A}lvarez-Mu{\~n}iz}}},
  \bibinfo{author}{\bibfnamefont{W.~R.} \bibnamefont{{Carvalho}}},
  \bibnamefont{and} \bibinfo{author}{\bibfnamefont{E.}~\bibnamefont{{Zas}}},
  \bibinfo{journal}{\ap} \textbf{\bibinfo{volume}{35}}, \bibinfo{pages}{325}
  (\bibinfo{year}{2012}), \eprint{1107.1189}.

\bibitem[{\citenamefont{{Corstanje} et~al.}(2023)}]{2023arXiv230613514C}
\bibinfo{author}{\bibfnamefont{A.}~\bibnamefont{{Corstanje}}}
  \bibnamefont{et~al.}, \bibinfo{journal}{arXiv e-prints}
  \bibinfo{eid}{arXiv:2306.13514} (\bibinfo{year}{2023}), \eprint{2306.13514}.

\bibitem[{\citenamefont{Decoene}(2020)}]{decoene:tel-03153273}
\bibinfo{author}{\bibfnamefont{V.}~\bibnamefont{Decoene}},
  \bibinfo{type}{Theses}, \bibinfo{school}{{Sorbonne Universit{\'e}}}
  (\bibinfo{year}{2020}),
  \urlprefix\url{https://theses.hal.science/tel-03153273}.

\bibitem[{\citenamefont{Decoene et~al.}(2021)\citenamefont{Decoene,
  Martineau-Huynh, Tueros, and Chiche}}]{Decoene:2021jF}
\bibinfo{author}{\bibfnamefont{V.}~\bibnamefont{Decoene}},
  \bibinfo{author}{\bibfnamefont{O.}~\bibnamefont{Martineau-Huynh}},
  \bibinfo{author}{\bibfnamefont{M.~J.} \bibnamefont{Tueros}},
  \bibnamefont{and} \bibinfo{author}{\bibfnamefont{S.}~\bibnamefont{Chiche}},
  \bibinfo{journal}{PoS} \textbf{\bibinfo{volume}{ICRC2021}},
  \bibinfo{pages}{211} (\bibinfo{year}{2021}).

\bibitem[{\citenamefont{{Abbasi} et~al.}(2018)}]{2018ApJ...865...74A}
\bibinfo{author}{\bibfnamefont{R.~U.} \bibnamefont{{Abbasi}}}
  \bibnamefont{et~al.}, \bibinfo{journal}{\apj} \textbf{\bibinfo{volume}{865}},
  \bibinfo{eid}{74} (\bibinfo{year}{2018}), \eprint{1803.01288}.

\end{thebibliography}
% \begin{thebibliography}{99}
% \end{thebibliography}

%% Full authors list (ONLY FOR COLLABORATIONS)
% \clearpage
% \section*{Full Authors List: GRAND Collaboration}
% \input{authors.tex}
%
%\noindent \textbf{Note comment afterwards:} Collaborations have the possibility to provide an authors list in xml format which will be used while generating the DOI entries making the full authors list searchable in databases like Inspire HEP. \\
%
%\scriptsize
%\noindent
%first.author$^1$, 
%second.author$^2$, 
%third.author$^3$ % .... more names
%and 
%last.author$^{n}$ \\
%
%\noindent
%$^1$first.affiliation.
%$^2$second.affiliation. % .... more affiliation
%$^{m}$last.affiliation.

\end{document}